# Linear temperature dependence of electron spin resonance linewidths in $La_{0.7}Ca_{0.3}MnO_3$ and $YBaMn_2O_6$

D. L. Huber

Department of Physics, University of Wisconsin-Madison, Madison, WI 53706

Abstract

We analyze recent electron spin resonance (ESR) experiments in $La_{0.7}Ca_{0.3}MnO_3$ and $YBaMn_2O_6$ focusing on the behavior of the linewidth at high temperatures where it is a linear function of the temperature. Noting that the *g*-factors of the resonances are characteristic of the $Mn^{4+}$ ion in a cubic environment, we make the assumption that the linewidth involves the static susceptibility of the $Mn^{4+}$ spins which we analyze in the molecular field approximation. We conclude that the linear dependence on temperature is associated with the susceptibility having a Curie or Curie-Weiss form while the temperature-dependent relaxation mechanism has a microscopic rate proportional to the temperature. In $La_{0.7}Ca_{0.3}MnO_3$, the $Mn^{4+}$ susceptibility has the ferromagnetic Curie-Weiss form, and the static contribution to the linewidth arising from distortions of the oxygen octahedra is absent due to motional narrowing brought on by the rapid hopping of the $e_g$ polarons. In $YBaMn_2O_6$ either of two scenarios is possible. The $Mn^{4+}$ susceptibility above 520 K is Curie-like and the static term is present, or the susceptibility has the antiferromagnetic Curie-Weiss form and the static term is absent due to motional narrowing. It is concluded that the Curie model, with offsetting double exchange and and superexchange Curie-Weiss parameters, is the more likely scenario. It is suggested that the linear-*T* variation of the linewidth in both materials arises from either a Korringa-like mechanism involving interactions with mobile carriers or from a spin-phonon process coming from interactions between the $Mn^{4+}$ ions and the lattice vibrations.

E-mail address: huber@src.wisc.edu

Physics Department, University of Wisconsin-Madison, 1150 University Avenue, Madison, WI 53706

1. Introduction

Recent high-temperature electron spin resonance (ESR) studies of manganite materials have revealed systems where the linewidth, $\Delta H(T)$, is a linear function of the temperature [1,2], i.e. $\Delta H(T) = \delta + \beta T$. In [1] experiments were reported for the widely studied manganite $La_{1-x}Ca_xMnO_3$ for $0 \leq x \leq 0.7$ over the temperature range 300 K $\leq T \leq$ 580 K, the upper limit of the measurement. For $x = 0.3$, the linewidth of the crystalline sample is linear in the temperature, with $\delta = -570$ Oe and $\beta = 2.78$ Oe/K. In [2], the linewidth in A-site ordered $YBaMn_2O_6$ varies linearly with temperature from the structural transition at 520 K to 930 K, the upper limit of the measurement, with $\delta = 420$ Oe and $\beta = 1.07$ Oe/K. In both systems, the measured g-factor is close to the single-ion value for $Mn^{4+}$ in cubic insulators [1 - 4] indicating the resonance only involves the $Mn^{4+}$ spins.

The purpose of this note is to analyze the temperature dependence of the linewidth in the linear-$T$ regime. At high temperatures, the resonating Mn ions are in the 4+ state and the mobile $e_g$ electrons are charge carriers. When only the $Mn^{4+}$ ions are taking part in the resonance, the general equation for the linewidth must be adjusted. The appropriate expression takes the form

$$\Delta H(T) = (\chi_0^{Mn4+}(T) / \chi^{Mn4+}(T))(\Delta H_0 + BT), \qquad (1)$$

where the superscript Mn4+ in $\chi^{Mn4+}$ indicates that it is the susceptibility of the $Mn^{4+}$ array, not the total susceptibility of the combined system of $Mn^{4+}$ ions and $e_g$ electrons (or $Mn^{3+}$ ions), and $\chi_0^{Mn4+}$ denotes the corresponding Curie susceptibility. The $Mn^{4+}$ susceptibility, in this case, is defined by the ratio of the $Mn^{4+}$ magnetization induced by a *field that acts only on the $Mn^{4+}$ ions* to the magnitude of that field. Equation (1) comes with the implicit assumption that the $Mn^{4+}$ spins are not coherently coupled or 'bottlenecked' with the spins of the $e_g$ electrons – an issue we discuss in greater detail below. In the mean-field approximation, linear behavior in the linewidth occurs when either $\chi^{Mn4+}$ has the Curie form and $B > 0$ or when $\chi^{Mn4+}$ has the Curie-Weiss form and $\Delta H_0 = 0$. At this point, it is useful to consider the two materials separately. We begin with $La_{0.7}Ca_{0.3}MnO_3$

2. $La_{0.7}Ca_{0.3}MnO_3$

Since the constant term, $\delta$, in $La_{0.7}Ca_{0.3}MnO_3$ is negative, it can not be assigned to anisotropic effects. We interpret the linewidth behavior in terms of a ferromagnetic Curie-Weiss approximation for $\chi^{Mn4+}$, $(T - \Theta)^{-1}$ with Curie-Weiss temperature $\Theta = -\delta/\beta = 205$ K. As noted above, with the Curie-Weiss form, strictly linear behavior is obtained only when $\Delta H_0 = 0$. We attribute the absence of the anisotropy term to 'motional narrowing'. Motional narrowing comes about when the fluctuating lattice distortions generated by the hopping of the $e_g$ polarons reduce or eliminate the static anisotropy field. The condition for motional narrowing is

$$g\mu_B H^{rms}_{aniso}\tau / \hbar \ll 1, \tag{2}$$

where $H^{rms}_{aniso}$ is the root mean square anisotropy field, given by the square root of the second moment of the lineshape function, and $\tau$ is the correlation time for the fluctuations in the lattice configuration which we identify with the reciprocal of the polaron hopping rate. For small polarons at high temperatures, $1/\tau \sim$ optical phonon frequency, and Eq. (2) is generally satisfied. We emphasize that motional narrowing is associated with large amplitude fluctuations in the anisotropy field and is not important when the carriers are weakly coupled to the lattice. It also should be mentioned that the fluctuations in the anisotropy field may have a non-zero mean value leading to a finite value for $\Delta H_0$. Since there is no evidence of a finite value in the data, we conclude that $\Delta H_0/BT \ll 1$ over the temperature range of the linear behavior.

The presence of the linear-$T$ term in the linewidth over a wide temperature range suggests a mechanism analogous to Korringa broadening seen in ESR studies of magnetic impurities in metals where the linewidth arises from the interaction of the impurity ion with the conduction electrons in a relative bandwidth $kT/E_f$ at the Fermi surface. However, measurements of the resistivity at high temperatures in $La_{1-x}Ca_xMnO_3$ [5] indicate thermally activated behavior rather than conventional band transport possibly calling into question the traditional Korringa interpretation. In addition to the Korringa mechanism, the spin-lattice interaction also gives rise to a linear temperature dependence in processes involving the creation or destruction of a single phonon. We have extended earlier calculations of the one-phonon contribution to the ESR linewidth [6] and shown [7] that the phonon contribution to the width is limited by exchange narrowing so that only those phonon modes whose frequency is less than a spin-spin relaxation rate, $\gamma^{max}$, associated with isotropic exchange interactions between the resonating ions, contribute to the linewidth. When $T > \gamma^{max}$, the band of contributing modes generates a linear-$T$ term in the linewidth, similar to Korringa broadening. In [7] we estimate $\gamma^{max} \approx 260$ K, which is less than the onset temperature for linear behavior, 300 K.

3. $YBaMn_2O_6$

As noted, in Ref. [2] the authors report that the linewidth in $YBaMn_2O_6$ varied as a positive constant plus a positive term proportional to the temperature. This variation is consistent with Eq. (1), either when the $Mn^{4+}$ susceptibility has the Curie form, $\chi^{Mn4+} \sim 1/T$ with both $\Delta H_0$ and $B$ positive or when it has the antiferromagnetic Curie-Weiss form, $\chi^{Mn4+} \sim (T - \Theta)^{-1}$, with $\Theta = -\delta/\beta = -395$ K and $\Delta H_0 = 0$ (presumably due to motional narrowing). Measurements of the total susceptibility of $YBaMn_2O_6$ in the high temperature phase show ferromagnetic Curie-Weiss behavior with $\Theta = 286$ K [8]. As in $La_{1-x}Ca_xMnO_3$, the ferromagnetic behavior arises from the double exchange mechanism. Curie-like behavior of the $Mn^{4+}$ susceptibility could result from a near-cancellation of interactions between the $Mn^{4+}$ ions due to the presence of an effective ferromagnetic coupling along with the antiferromagnetic superexchange coupling [9]. In the Curie-Weiss approximation, this happens when $\Theta = |\Theta_{DE} + \Theta_{AFM}|/T << 1$, where $\Theta_{DE}$ is the effective Curie-Weiss temperature associated with the double exchange interaction, and $\Theta_{AFM}$ is the antiferromagnetic Curie-Weiss temperature.

In [2], results are presented for the integrated intensity of the ESR signal. It was found that the integrated intensity showed ferromagnetic Curie-Weiss like behavior with $\Theta = 387$ K. Although it would appear that this finding contradicts our analysis of $\chi^{Mn4+}$, this is not the case. As noted in the Introduction, $\chi^{Mn4+}$ characterizes the static response of the $Mn^{4+}$ array to an applied field that acts only on the $Mn^{4+}$ ions, whereas the integrated intensity characterizes the response of the total system to a field that acts on both the $Mn^{4+}$ ions and the carriers. The ferromagnetic behavior seen in the integrated intensity suggests that the antiferromagnetic Curie-Weiss model with motional narrowing is a less plausible interpretation than the Curie model with offsetting double exchange and superexchange Curie-Weiss temperatures.

The Korringa interpretation of the linewidth in this system is plausible in view of the fact that the linear-T behavior is only observed above 520 K where the system is in a metallic phase [2]. For the phonon interpretation of the temperature dependence of the ESR linewidth in $YBaMn_2O_6$, a critical question is whether the temperature is high enough to satisfy the condition $T > \gamma^{max}$ over the range of the experiment. In [7] we estimate $\gamma^{max} \approx 300$ K, which is well below the onset of linear-$T$ behavior, suggesting that phonons may also be contributing to the linewidth

4. Discussion

We begin by pointing out that the linear behavior of the ESR linewidth at high temperatures has also been seen in the compounds $ReBaMn_2O_6$ with Re = Dy, Tb, Gd, Eu and Sm, as well as in $Nd_{0.67}Ca_{0.33}MnO$ [10], and in earlier ESR measurements on $La_{1-x}Sr_xMnO_3$ with $x = 0.2$ and 0.3 [11].

In reviewing our analysis of the linewidths, we emphasize there are important questions requiring further investigation. A critical issue is the behavior of $\chi^{Mn4+}$ in the presence of competing antiferromagnetic superexchange and ferromagnetic double exchange interactions. In $La_{0.7}Ca_{0.3}MnO_3$, ferromagnetic interactions dominate while in the case of $YBaMn_2O_6$, the mean field analysis of $\chi^{Mn4+}$ suggests that antiferromagnetic and ferromagnetic interactions are most likely of comparable importance.

In the Introduction, it was pointed out that Eq. (1) carries with it the implicit assumption that the $Mn^{4+}$ spins are not coherently coupled to the spins of the $e_g$ electrons as occurs in a 'bottleneck' situation where the total spin ($Mn^{4+}$ ions plus $e_g$ electrons) is an (approximate) constant of the motion [12]. The absence of a bottleneck is consistent with the appearance of the linear-$T$ term in the linewidth which is evidence of a direct coupling to the thermal bath. In view of its importance, the origin of the linear-$T$ term and its connection with double and superexchange interactions merit further study.

Acknowledgments

We would like to thank M. Auslender, J. Deisenhofer, and E. Rozenberg for helpful comments.